\documentclass[12pt,english]{paper}
\usepackage[T1]{fontenc}
\usepackage[latin1]{inputenc}
\usepackage{geometry}
\usepackage{graphicx}
\usepackage{setspace}
\usepackage[numbers]{natbib}

\makeatletter

 \newcommand{\lyxaddress}[1]{
   \par {\raggedright #1 
   \vspace{1.4em}
   \noindent\par}
 }
 \newenvironment{lyxlist}[1]
   {\begin{list}{}
     {\settowidth{\labelwidth}{#1}
      \setlength{\leftmargin}{\labelwidth}
      \addtolength{\leftmargin}{\labelsep}
      }}
   {\end{list}}

\usepackage{babel}
\makeatother
\begin{document}

\title{Direct demonstration of decoupling of spin and charge currents in
nanostructures}

\author{M. Urech, V. Korenivski, N. Poli, \& D. B. Haviland}

\maketitle

\lyxaddress{Nanostructure Physics, Royal Institute of Technology, 10691 Stockholm,
Sweden}

\begin{abstract}
The notion of decoupling of spin and charge currents is one of the
basic principles underlying the rapidly expanding field of Spintronics.
However, no direct demonstration of the phenomenon exists. We report
a novel measurement, in which a non-equilibrium spin population is
created by a point-like injection of current from a ferromagnet across
a tunnel barrier into a one dimensional spin channel, and detected
differentially by a pair of ferromagnetic electrodes placed symmetrically
about the injection point. We demonstrate that the spin current is
strictly isotropic about the injection point and, therefore, completely
decoupled from the uni-directional charge current.\newpage

\end{abstract}
Each individual electron in a conductive material transports a quantum
of charge as well as a quantum of spin. This charge-spin coupling,
fundamental for individual electrons, is generally broken for ensembles
of electrons. For the ensemble, spin is not conserved because spin-flip
scattering provides a means of reversing the spin of the electron.
No comparable mechanism exists for changing the charge of the electron,
and charge is strictly conserved. Even though the notion of spin-charge
decoupling of currents \cite{johnson85} is one of the basic concepts
underlying the rapidly expanding field of Spintronics \cite{Zutic},
no direct demonstration of the phenomenon exists and controversies
persist \cite{jedema01,johnson02,jedema02,johnson03,jedema03}. We
report a novel measurement, in which a non-equilibrium spin population
is created by a point-like injection of current from a ferromagnet
across a tunnel barrier into a metallic nanowire, and detected differentially
by a pair of ferromagnetic electrodes placed symmetrically about the
injection point. We demonstrate that the spin current in the wire
is strictly isotropic about the injection point and therefore completely
decoupled from the uni-directional charge current. This device creates
an ideal source of pure spin current that can be easily controlled
by the geometry of the device. This general concept of 'spin pumping'
is currently of great fundamental and technological interest, forming
the base for new types of spintronic devices \cite{sharma}. 

Fundamental to the rapidly growing field of Spintronics is the ability
to manipulate spin currents in nanostructures, which promises new
implementations of memory and logic \cite{Zutic}. The first suggestions
on spin injection and its detection date back a few decades \cite{aronov,johnson85}.
A spin polarized current, \[
j_{s}\equiv j_{\uparrow}-j_{\downarrow}\equiv\gamma(j_{\uparrow}+j_{\downarrow})\equiv\gamma j_{q},\]
 injected form a ferromagnet (F) into a nonmagnetic conductor (N)
creates a spin splitting of the chemical potential at the F/N interface,
which is directly determined by the spin polarization of the injected
current, $\gamma$, \[
\bigtriangledown\delta\mu\equiv\bigtriangledown(\mu_{\uparrow}-\mu_{\downarrow})=e\rho_{N}\mathbf{j}_{s}=e\rho_{N}\gamma\mathbf{j}_{q}.\]
 Here $\rho_{N}$ is the resistivity of N and $e$ is the electron
charge ($j_{s}$ and $j_{q}$ are collinear at the injection interface).
This non-equilibrium spin population decays away from the injection
point into N over a characteristic spin flip length $\lambda_{sf}$,
and is governed by a diffusion equation \begin{equation}
\bigtriangledown^{2}\delta\mu(\mathbf{r})=\frac{1}{\lambda_{sf}^{2}}\delta\mu(\mathbf{r}).\label{diff}\end{equation}
 It is essential to realise that the spin and charge currents are
generally decoupled everywhere in N except at the injection point.
This is easily illustrated by writing out the spin-up and spin-down
currents in N using $\mu_{\uparrow,\downarrow}(\mathbf{r})\equiv\mu_{0}\pm\frac{1}{2}\delta\mu(\mathbf{r})$,
\begin{equation}
\mathbf{j}_{\uparrow,\downarrow}(\mathbf{r})=\frac{1}{e\rho_{N}}\bigtriangledown\mu_{0}(\mathbf{r})\pm\frac{1}{2e\rho_{N}}\bigtriangledown\delta\mu(\mathbf{r}),\label{jupdown}\end{equation}
 with $\mu_{0}$ being the equilibrium, spin independent electrochemical
potential in N. $\mu_{0}$ is either constant or varies linearly with
coordinate in regions of non-zero electric field $E$, where it produces
a charge current, \[
\mathbf{j}_{q}=\frac{1}{e\rho_{N}}\bigtriangledown\mu_{0}=\sigma_{N}\mathbf{E}.\]
The important prediction of Eq.\ref{jupdown} for regions of zero
$E$ (outside the path of $j_{q}$) is that the current of spin-up
electrons is exactly compensated by the opposing current of spin-down
electrons, which diminishes the charge flow and doubles the spin flow
associated with $\delta\mu$: $\mathbf{j}_{\uparrow}=-\mathbf{j}_{\downarrow},\ \ \left|\mathbf{j}_{\uparrow}\right|=\left|\mathbf{j}_{\downarrow}\right|$.
For regions of N with non-zero $E$ the spin and charge currents are
superposed. The two currents are restricted to the same path only
in the pure 1-D case, typical for Giant Magneto Resistance structures
\cite{van-son,v-f}, but have very different profiles in the general
case where the driving electric field is non-uniform in N. To date
no experimental evidence exists, which unambiguously verifies spin-charge
current decoupling in nanostructures and controversies persist in
the analysis of multi-terminal spin based devices \cite{jedema01,johnson02,jedema02,johnson03,jedema03}.
Here we directly demonstrate that spin and charge currents are strictly
decoupled in diffusive 1D spin channels. 

Johnson and Byers propose an ideal system for demonstrating decoupling
of spin and charge currents \cite{johnson03}, consisting of a conductive
nano-wire with a spin imbalance created by a point injection of a
spin-polarized current into the wire and spin-sensitive detectors
placed symmetrically about the injection point, which is illustrated
in Fig. 1. The fundamental decoupling of the spin and charge currents
can be demonstrated by detecting spin signals of \emph{precisely equal
strength} in both directions from the injection point while the charge
current flows in only one direction. It is essential that the magnetic
detectors are closely spaced from the injector within the spin relaxation
length in N, $\sim$100 nm for typical metal films. The detectors
are probe tunnel junctions, which should have identical properties.
The magnetization states of the detectors must also be arranged by
external means to achieve parallel and anti-parallel configurations
with respect to the injector. For this 1-D geometry, with a point-like
injection at $x=0$, the solution of Eq.\ref{diff} is \begin{equation}
\delta\mu(x)\propto e^{-|x|/\lambda_{sf}}.\label{exp}\end{equation}
The spin signal at the detector is a direct measure of $\delta\mu$:
\begin{equation}
V_{s}=\pm\frac{\gamma}{2e}\delta\mu.\label{Vs}\end{equation}
This spin signal is compared for the local and non-local case, where
the detector is inside and outside the charge current path, respectively.
In this letter we report a simultaneous and differential measurement
of such local and non-local spin signals, which clearly demonstrates
a complete spin-charge decoupling for currents in nanostructures.

Fig. 2 shows a SEM micrograph of the sample fabricated using a two-angle
deposition through a lift-off mask, which was patterned by e-beam
lithography. The injector and detector junctions are formed by depositing
and oxidizing a $\sim$100 nm wide and 20 nm thick Al strip with a
subsequent deposition of 55, 65, and 50 nm wide and 40 nm thick Co
electrodes, overlapping the Al strip. The width and thickness of the
Al strip are much smaller than the spin flip length, making it a 1D
channel for spin diffusion. The width of the Co electrodes was varied
to achieve different switching fields and thereby multiple stable
magnetic states of the device. It is essential that the magnetic electrodes
switch in a single domain fashion. We found \cite{urech} that in
such Co based overlap junctions two well defined longitudinal magnetic
states with sharp inter-state transitions are obtained for narrow
electrodes, 90 nm and less in width. This length is comparable or
slightly smaller than the characteristic domain wall size in thin
Co. Thin, long, and narrow electrodes can be modelled as single domain
particles of uniaxial anisotropy, where the width/length ratio determines
the longitudinal switching field. Therefore, the smaller the width
the higher the switching field. Such a relation for our geometry is
also supported by exact micromagnetic simulations \cite{urech}. In
order to fine-tune the width of the Co electrodes we have adjusted
the layout of the devices such that the total dose during the mask
exposure, including the backscattered electrons, was slightly different
for the three magnetic electrodes. For example, backscattering is
strongest for the center electrode, making it the widest (65 nm) of
the three and therefore with the lowest switching field (1.13 kOe).
The outer Co electrodes were 50 and 55 nm wide with the switching
fields of 1.62 and 1.45 kOe, respectively.

The measurements have been done using a standard lock-in technique
at 7 Hz, which we found necessary for achieving $\sim0.1$\% level
resolution in the spin signal measurements. The ac technique resulted
in a weak but non-zero inductive crosstalk between the injection and
detection circuits, which scaled to zero with decreasing the excitation
frequency. This small constant background of $\sim30$ nV had no influence
on the spin-dependent signals studied, which was verified by measuring
the latter at different frequencies as well as a dc technique. The
external field was applied along the length of the Co electrodes.
The sample was electrically characterized as follows: injecting the
current from probe 3 to probe 4 ($I_{34}$), we measure the non-local
voltages $V_{10}$ and $V_{20}$ at two distances from the injection
point. In this non-local measurement configuration, the voltage probes
are outside the current path. According to Eqs. \ref{exp},\ref{Vs}
the ratio of these voltages for our 1-D geometry is \begin{equation}
V_{10}/V_{20}=\exp(d_{12}/\lambda_{sf}),\label{Vratio}\end{equation}
where $d_{12}=300$ nm is the electrode separation and $\lambda_{sf}=\sqrt{D\tau_{sf}}$
is the spin diffusion length, found to be 850 nm at 4 K. Here $D$
and $\tau_{sf}$ are the diffusion constant and spin relaxation time,
respectively. The diffusion constant was calculated from the measured
resistivity of the wire, $\rho=4.0~\mu\Omega$cm, using the Einstein
relation $D=[e^{2}\rho N(E_{F})]^{-1}=65$ cm$^{2}/$s, where $e$
is the electron charge and $N(E_{F})=2.4\times10^{28}$ eV$^{-1}$m$^{-3}$
is the density of states of Al at the Fermi energy \cite{Papa}. The
resulting spin relaxation time is $\tau_{sf}=110$ ps, much longer
than the Drude scattering time, $\tau_{p}/\tau_{sf}\sim10^{-4}$.
From the measured spin voltages we can also determine the effective
spin polarization $P=12\%$ ($V_{i0}\propto P^{2}$ \cite{johnson03}).
These values are in good agreement with the recent spin injection
studies on Al films \cite{jedema02,Valenzuela} but differ from the
early single crystal Al data \cite{johnson85,Lubzens} due to stronger
impurity and surface scattering in thin films.

In Fig. 3a the local spin signal ($V_{30}$, solid line) and the non
local spin signal ($V_{10}$, dashed line) are plotted as a function
of the applied magnetic field, for $I_{42}$=5 $\mu$A. Isotropic
spin diffusion necessarily results in these two spin voltages being
identical in magnitude. A precondition is that the detectors have
the same physical characteristics, such as polarization, which is
achieved in our case by the simultaneous deposition of all three junctions.
The switching of the injector electrode at 1.13 kOe ($\downarrow\downarrow\downarrow\ \rightarrow\ \downarrow\uparrow\downarrow$)
reverses the sign of the injected spin population in the wire. This
is measured simultaneously by the two detectors and yields \emph{exactly}
the same spin voltages, thus demonstrating that the spin currents
are identical in both directions while the charge current flows only
in one direction. As the field is increased, detector 1 switches ($\downarrow\uparrow\downarrow\ \rightarrow\ \uparrow\uparrow\downarrow$),
followed by a switching of the second detector ($\uparrow\uparrow\downarrow\ \rightarrow\ \uparrow\uparrow\uparrow$),
again demonstrating the symmetry in the spin current. 

An even more direct demonstration of the effect is the differential
spin signal ($V_{13}$) shown in Fig. 3b, which, in addition to eliminating
the fabrication related uncertainties, also rules out possible measurement
errors, such as small differences in gain and drift of the separate
amplifiers and small spin misalignments in the junction areas. The
raw differential signal $V_{13}$ contains the resistive, spin-\emph{independent}
contribution from $V_{23}$ ($V_{23}/I_{42}=6\ \Omega$) as well as
both spin-\emph{dependent} signals from $V_{30}$ and $V_{10}$. The
switching of the injector at 1.13 kOe ($\downarrow\downarrow\downarrow\ \rightarrow\ \downarrow\uparrow\downarrow$)
leaves the differential spin voltage unchanged to within the experimental
uncertainty of $<1\%$ (Fig. 3b), while the two component signals
change by the maximum amount (Fig.3 a). The slight skew in the signals
in Fig. 3a,b as the reversing field is increased is connected with
small spin perturbations in the junction areas caused by the opposing
external field (<1.65 kOe). We emphasize however, that the power of
our symmetric detection scheme is in eliminating this micromagnetic
uncertainty. The change of the magnetic state of the injector at a
given point in field (1.13 kOe) changes both spin signals by the same
maximum amount. However, the magnetic states of the detector electrodes,
even if slightly misaligned, are not affected by this instantaneous
switching of the injector. The differential signal therefore remains
unchanged (Fig. 3b at 1.13 kOe).

A difference in width for the two detector electrodes was necessary
to separately control the respective magnetic states by an external
magnetic field. The small uncertainty this introduces can be estimated
as follows. From Eqs. \ref{exp},\ref{Vs},\ref{Vratio} the variation
in detector width from 50 to 55 nm can be estimated to produce a 0.06\%
change in the detector voltage. A change in the detector width of
even a factor of 2 (50 nm to 100 nm) would still have a very small
effect on the the effective spin signal detected ($\sim0.1$\%). The
uncertainty due to the difference in width of our two detectors (5
nm) is thus negligible, determined by the relatively large scale set
by the spin diffusion length. The relative weight of the detector
placement error is on the other hand larger, and follows directly
from Eq. \ref{Vratio}. Our experimental placement error for the sample
discussed above is approximately 5 nm corresponding to a 0.6\% change
in the detector voltage, which is still within the $\sim1$\% precision
we report. We have measured several samples with well controlled magnetic
switching and similar geometrical uncertainties, all confirming the
above result at the 1\% level.

In summary, we use a multi-terminal spintronic device to unambiguously
verify the complete decoupling of spin and charge currents in the
regime of diffusive transport. The important consequence of this decoupling
is that numerous recently proposed multi-terminal spin-based devices
should benefit in sensitivity by use of non-local as opposed to local
spin currents.

\section*{Acknowledgements}

Financial support from the Swedish Foundation for Strategic Research
(SSF) is gratefully acknowledged.

\newpage
\section*{Figure captions:}

\begin{lyxlist}{00.00.0000}
\item [Fig.1]Schematic of the mechanism of spin and charge transfer for
a current injected from a ferromagnet (electrode {}``2'') into a
non-magnetic conducting wire. The spin current is symmetric about
the injection point and is due to opposing currents of spin-up and
spin-down electrons (see Eq. 1). The density of the spin current ($J_{s}$,
grey shaded) decays exponentially from the injection point. The charge
current is asymmetric and unpolarised, ($J_{q}$, white). The density
of the charge current is uniform between the injector (Inj) and ground
(G). The injected non-equilibrium spin accumulation can be sensed
by ferromagnetic detectors in the vicinity of the injection point
(electrodes {}``1'' and {}``3'') either with respect to the floating
end of the nano-wire (F) or differentially. 
\item [Fig.2]SEM micrograph of the sample consisting of an oxidized 100
nm wide and 15 nm thick Al strip overlapped by three 40 nm thick Co
electrodes of 55, 65, and 50 nm in width. The junction resistances
are $\sim20\  k\Omega$. The injection-detection configuration for
non-local and differential (local versus non-local) measurement configurations
is indicated.
\item [Fig.3]a) local (solid) and non-local (dashed) spin signals measured
simultaneously as a function of external field. b) differential signal
as a function of field. The constant, spin-\emph{independent} resistive
offset ($V_{23}/I_{42}=6\ \Omega$) has been subtracted from the raw
$V_{30}$ and $V_{13}$ signals, allowing a direct comparison of the
spin-dependent signals. The behavior in negative fields is identical.
The $\sim30$ nV offset in $V_{10},V_{30}$ is due to a spin-independent
inductive crosstalk (see text). All data were taken at 4 K with $I_{42}=5\ \mu$A.
The arrows indicate the magnetization states of the three ferromagnetic
electrodes.\newpage

\end{lyxlist}

\begin{figure}
\centering
\includegraphics[%
  width=1.0\textwidth]{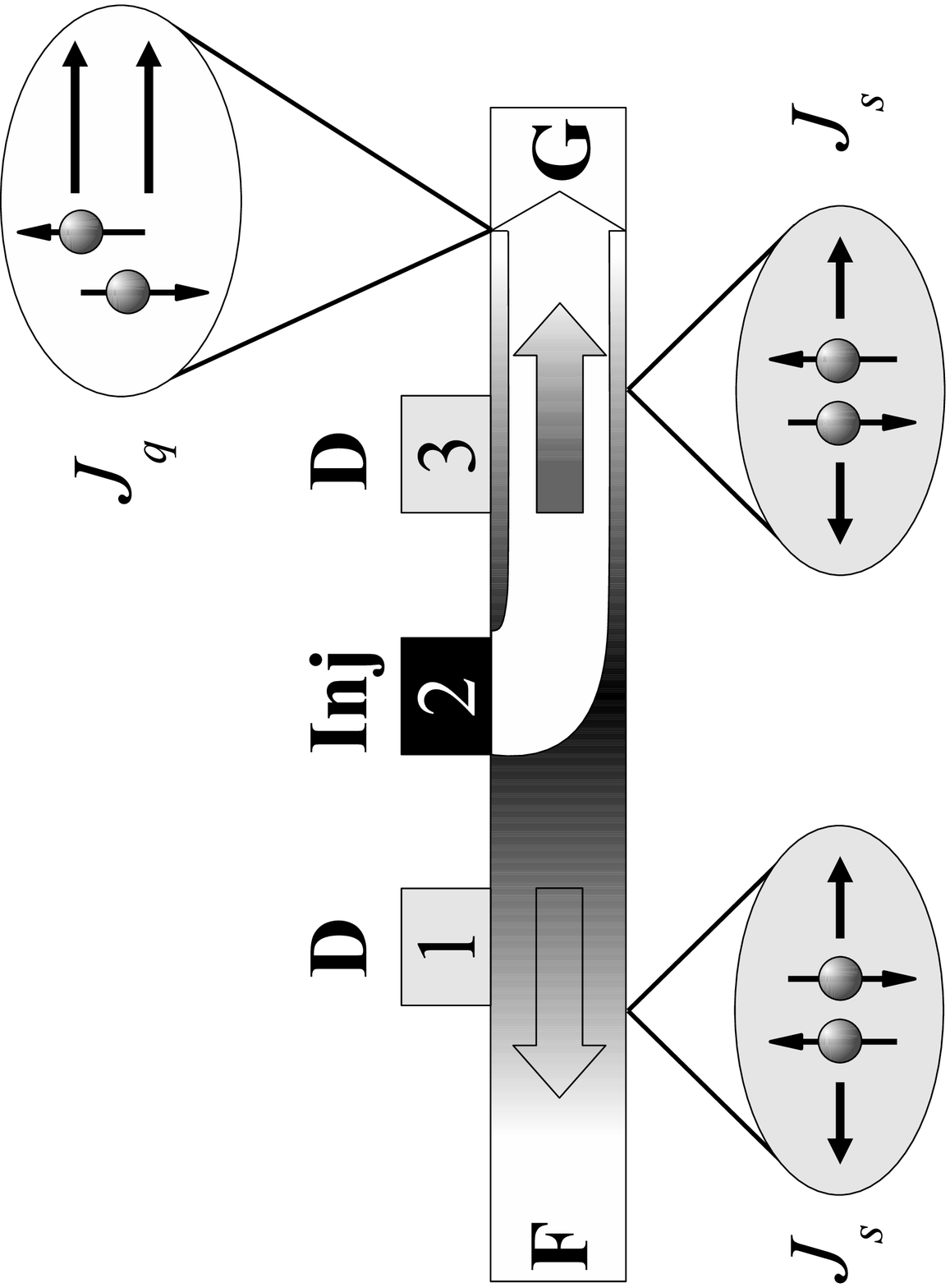}
\caption{Urech \emph{et al.} }
\end{figure}

\begin{figure}
\centering
\includegraphics[%
  width=1.0\textwidth]{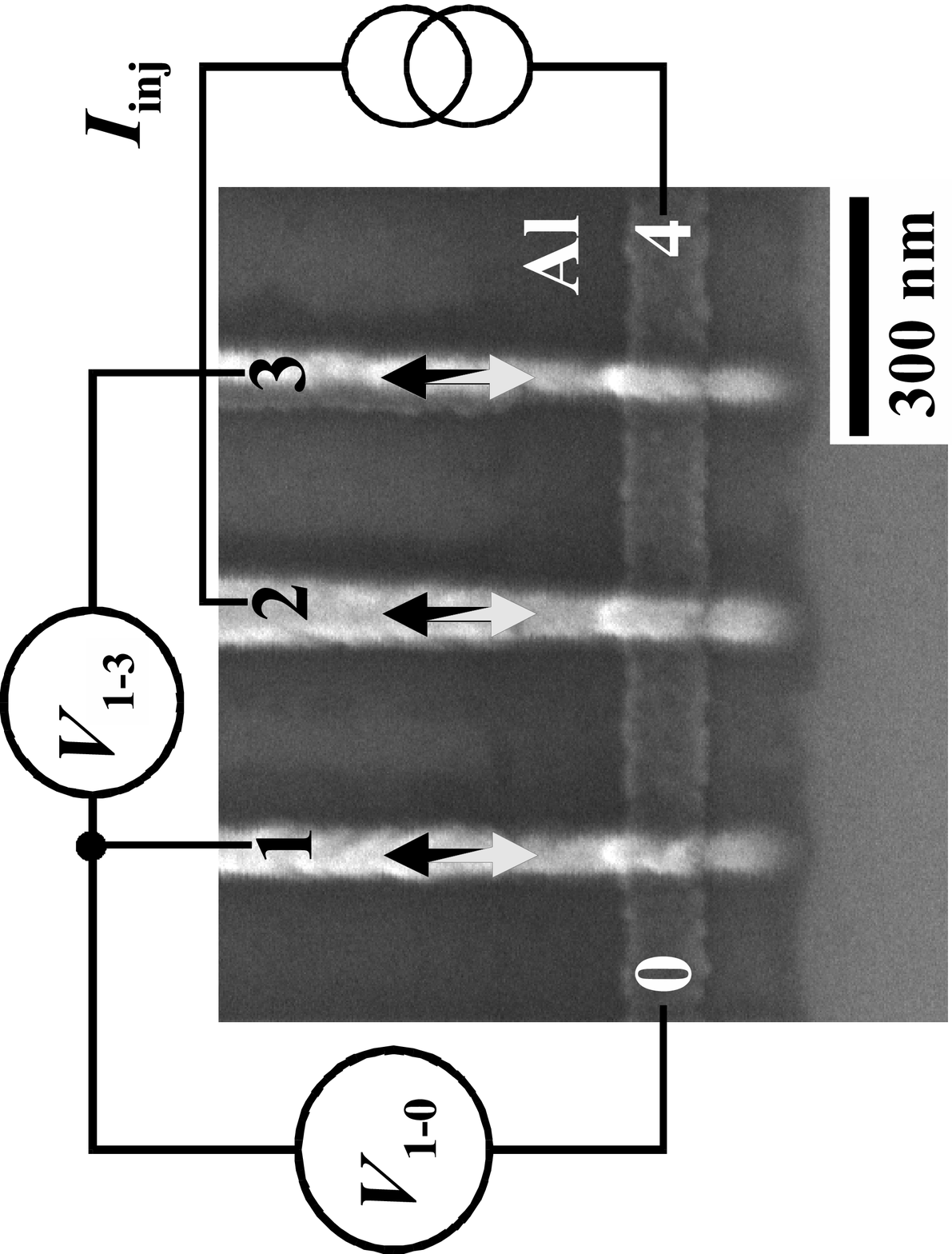}
\caption{Urech \emph{et al.}}
\end{figure}

\begin{figure}
\centering
\includegraphics[%
  width=1.0\textwidth]{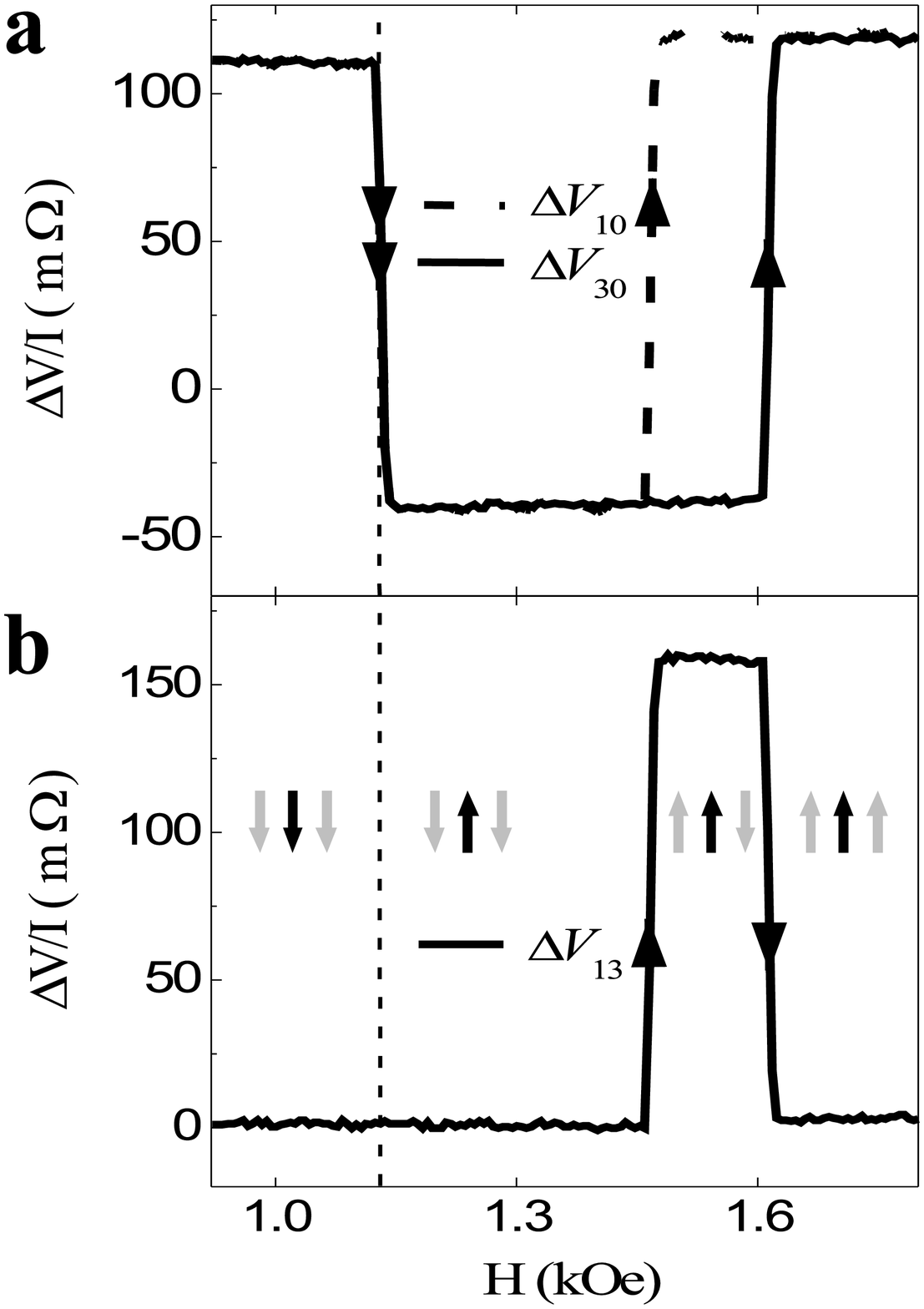}
\caption{Urech \emph{et al.}}
\end{figure}

\end{document}